# Cyber Threats to Canadian Federal Election: Emerging Threats, Assessment, and Mitigation Strategies

Nazmul Islam[1,*], Soomin Kim[2], Mohammad Pirooz[2], and Sasha Shvetsov[2]
[1]School of Computing, Queen's University, Kingston, Canada
[2]Royal Military College of Canada, Kingston, Canada
*nazmul.islam@queensu.ca (Corresponding Author)

*Abstract*— As Canada prepares for the 2025 federal election, ensuring the integrity and security of the electoral process against cyber threats is crucial. Recent foreign interference in elections globally highlight the increasing sophistication of adversaries in exploiting technical and human vulnerabilities. Such vulnerabilities also exist in Canada's electoral system that relies on a complex network of IT systems, vendors, and personnel. To mitigate these vulnerabilities, a threat assessment is crucial to identify emerging threats, develop incident response capabilities, and build public trust and resilience against cyber threats. Therefore, this paper presents a comprehensive national cyber threat assessment, following the NIST Special Publication 800-30 framework, focusing on identifying and mitigating cybersecurity risks to the upcoming 2025 Canadian federal election. The research identifies three major threats: misinformation, disinformation, and malinformation (MDM) campaigns; attacks on critical infrastructure and election support systems; and espionage by malicious actors. Through detailed analysis, the assessment offers insights into the capabilities, intent, and potential impact of these threats. The paper also discusses emerging technologies and their influence on election security and proposes a multi-faceted approach to risk mitigation ahead of the election.

## I. INTRODUCTION

The integrity of electoral processes is increasingly dependent on effective cybersecurity measures in the modern digital age. As elections become more reliant on technology for various functions such as election campaigns, voter registration, vote tallying, and information dissemination, they also become more vulnerable to cyber threats by foreign actors due to the ability to transcend geographic limits. The past decade has witnessed significant instances of foreign interference in elections across various countries, highlighting the increasing sophistication and persistence of adversaries. In 2016, Russia conducted a comprehensive campaign to influence the U.S. presidential election through social media manipulation, deploying bots and trolls, and hacking to leak sensitive information [1], [2]. Similarly, during the Brexit referendum in the United Kingdom in 2016, Russian actors sought to sway public opinion by promoting anti-EU sentiment and deepening societal divisions[1], [2]. In 2017, Germany's Bundestag election was targeted by Russian disinformation campaigns designed to amplify existing societal tensions [3]. France also experienced cyberattacks during its 2017 presidential election when President Macron's campaign was compromised through leaked emails and documents [4]. The European Parliament elections in 2019 saw multiple EU member states reporting misinformation and disinformation attacks from unattributed sources [5]. In 2020, Iran interfered in the U.S. presidential election by using voter information to send threatening emails[6]. Most recently, in 2023, Taiwan's elections were disrupted by China's use of deepfake technology to create societal division [7], [8]. The Canada Centre for Cyber Security has reported that in 2022, just over a quarter of national elections worldwide experienced at least one documented cyber incident of which 25% were NATO member countries and 35% OECD countries [9]. This marks a significant increase in cyber threat activity targeting elections globally, rising from 10% of national elections in 2015 to 26% in 2022 as shown in Figure 1. Drawing a linear graph from the average of each year, this percentage is expected to increase over 40% by 2028.

In recent years, Canada has faced increasing risks from cyber threats, primarily attributed to foreign actors such as Russia and China, who aim to influence electoral outcomes and voter perceptions through hacking and disinformation campaigns [10]. This was particularly evident during the 2021 general election, where reports suggested attempts by Chinese officials and state media to influence voters of Chinese origin against the Conservative Party, although this did not significantly impact the election outcome. The Canadian electoral process, managed by Elections Canada, is designed to be non-partisan and transparent, relying on traditional paper ballots to mitigate risks associated with electronic voting systems. However, vulnerabilities remain, particularly in the digital realm where political parties and election infrastructure can be targeted by cyberattacks like distributed denial-of-service (DDoS) attacks and ransomware. To safeguard its democracy, Canada has implemented proactive measures such as the Elections Modernization Act, which introduces spending limits and increases transparency [11]. Additionally, the establishment of the Canadian Centre for Cyber Security and the implementation of the Critical Election Incident Public Protocol are key steps taken to safeguard elections [12], [13]. These initiatives involve monitoring potential threats, offering cybersecurity guidance to political parties and administrators, enhancing communication and information sharing of political concerns, and ensuring transparency of incidents that may undermine election integrity.

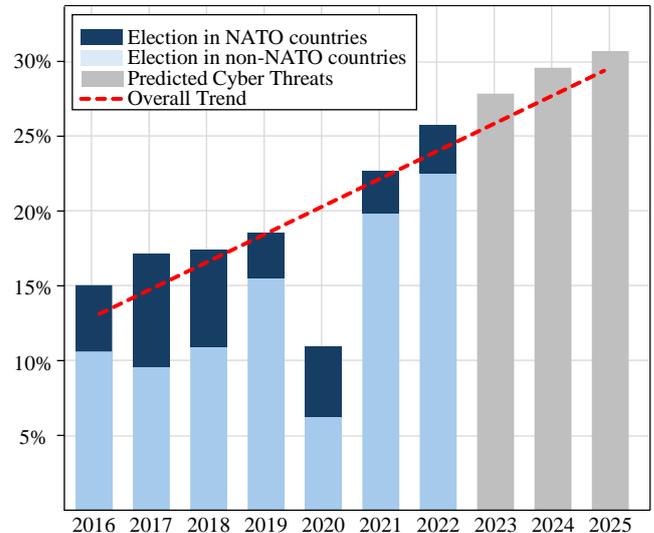

Fig. 1. Cyber threats to national-level elections. Adopted from [9].

While Canada's hybrid approach of combining traditional paper ballots with digital systems (for voter registration and results compilation) reduces some risks associated with electronic voting machines, vulnerabilities persist in areas such as voter registration databases, communication networks, and political party systems. These components are attractive targets for cybercriminals seeking to disrupt elections or influence political outcomes. Additionally, the increase in misinformation on social media and the rise of generative artificial intelligence (AI) [14] present significant threats to democratic integrity, as false narratives can rapidly spread and manipulate public opinion [15], [16]. As a result, robust cybersecurity measures are essential not only to safeguard the technical integrity of electoral systems and prevent operational disruptions but also to maintain public trust in democratic institutions by securing sensitive data and combating the spread of misinformation and interference in public discourse. This paper aims to provide a comprehensive assessment of cyber threats facing Canada's 2025 federal election using the NIST Special Publication 800-30 framework [17]. The main contributions include (1) Identifying emerging threats specific to Canada's election, (2) Analyzing potential impacts on election integrity through a detailed diamond model analysis and threat assessment, and (3) Proposing a multi-faceted risk mitigation strategy.

The rest of the paper is organized as follows: section. II covers cybersecurity standards and best practices for elections, section. III discusses the emerging threat landscape, section. IV addresses the risk assessment of these threats, section. V presents mitigation strategies, and finally, the conclusion summarizes the findings.

## II. EXISTING CYBERSECURITY STANDARDS AND PRACTICES

To ensure the integrity and security of elections, it is crucial to adopt comprehensive cybersecurity standards and best practices. These guidelines help protect election infrastructure from various cyber threats, ensuring a fair and transparent electoral process. Industry-specific guidelines, such as those provided by the election infrastructure information sharing and analysis center (EI-ISAC), offer resources tailored to the unique needs of election officials [18]. EI-ISAC provides threat intelligence, incident response support, and cybersecurity training to enhance the resilience of election systems against cyber threats.

The NIST Cybersecurity Framework provides a structured approach to managing cybersecurity risks through its core functions: Identify, Protect, Detect, Respond, and Recover [19]. This framework helps organizations prioritize their cybersecurity efforts based on risk assessments and aligns with objectives. Complementing these efforts, the Canadian centre for cyber security (CCCS) offers guidance to election authorities on protecting electoral processes from cyber threats [20]. The CCCS emphasizes the importance of safeguarding network integrity, ensuring data confidentiality, and maintaining system availability while providing resources for detecting and responding to incidents effectively.

International best practices and frameworks such as the ISO/IEC 27000 series, particularly ISO/IEC 27032, focuses on cybersecurity guidelines for protecting information in cyberspace by addressing risks related to online activities [21]. Additionally, the CIS Critical Security Controls offer a prioritized set of actions to improve cybersecurity posture by addressing common vulnerabilities like poor configuration management and outdated software [22]. Key measures like implementing robust access control mechanisms ensure that only authorized personnel can access sensitive election data. Encryption further protects this data, both during transmission and at rest, from unauthorized access. Regular backups are vital for ensuring data recovery in the event of cyber incidents or system failures. Additionally, conducting regular audits helps identify potential vulnerabilities and ensures compliance with established security policies.

Canada has a variety of measures to protect its electoral process, emphasizing the enhancement of cybersecurity across critical systems and ensuring election integrity. The critical cyber systems protection act (CCSPA) establishes a regulatory framework to enhance cybersecurity for critical systems in sectors such as telecommunications, energy, and transportation [23]. It mandates designated operators to report cybersecurity incidents, develop cybersecurity plans, and comply with directives aimed at mitigating identified threats. Amendments to the Telecommunications Act empower the Canadian government to enforce security measures within telecommunications networks. These measures include prohibiting certain suppliers' products and ensuring operators secure their systems against potential threats. Furthermore, Elections Canada has taken proactive steps to secure the electoral process. This includes collaboration with security agencies like the communications security establishment (CSE) and the Canadian centre for cyber security to monitor emerging threats and enhance situational awareness. New IT systems are developed with security as a core element, adhering to stringent government cybersecurity policies. Continuous training programs are provided to employees on safeguarding information and practicing good cyber hygiene. Additionally, the Critical Election Incident Public Protocol ensures clear communication with Canadians during election periods about incidents that could threaten election integrity [24]. The security intelligence threats to elections (SITE) task force collaborate across government agencies to monitor and address threats related to foreign interference [24].

The implementation of robust cybersecurity standards and best practices provides a critical foundation to maintaining a secure electoral environment in Canada by addressing both technological vulnerabilities and potential foreign interference threats. However, as we prepare for the upcoming federal election, it is important to proactively assess the evolving cyber threat landscape.

## III. EMERGING CYBER THREATS

The chapter delves into the emerging cyber threats to the upcoming Canadian federal election, offering a detailed analysis of the cyber threat landscape using the diamond model framework. The analysis identifies three major threats, providing a comprehensive understanding of the threat landscape by examining adversaries, capabilities, infrastructure, and potential victims for each of the threats.

*A. Misinformation, Disinformation, Malinformation (MDM)*

Misinformation, disinformation, and malinformation (MDM) are forms of false or misleading information that can cause harm. Misinformation is false information shared without harmful intent. Disinformation is deliberately false information created to mislead or harm. Malinformation is factual but used out of context to mislead or harm.

*Adversary*: The primary adversaries in MDM campaigns include foreign governments and political extremist groups. State-sponsored actors, particularly from countries like Russia and China, utilize information warfare to influence public opinion and electoral outcomes [25]. These actors aim to destabilize the Canadian political climate by creating divisions among political parties and undermining public trust in government institutions. Political extremist groups, both domestic and international, also contribute to this destabilization by spreading polarizing narratives that can incite unrest.

*Capability*: The capabilities employed by these adversaries are sophisticated and multifaceted. Social engineering techniques are used to manipulate individuals into divulging confidential information or spreading false narratives. Data exfiltration tactics allow adversaries to extract sensitive information from compromised systems, which can then be used to fuel disinformation campaigns. Additionally, campaigns designed to erode public trust in government institutions and the electoral process are a key strategy, leveraging false narratives to sow doubt and confusion among the electorate.

*Infrastructure*: The infrastructure targeted by MDM campaigns includes social media platforms, government databases, and campaign websites [26]. Social media platforms like Facebook and X are particularly vulnerable due to their ability to rapidly disseminate false information through algorithms and coordinated fake accounts. These platforms target specific demographics, such as the 25-34 age group [27], creating an illusion of widespread consensus on false narratives. Government databases containing sensitive voter information are also at risk of compromise, which can undermine trust in the electoral process. Campaign websites may be hacked or manipulated to spread false information, misleading voters about candidates or election procedures.

*Victims*: The primary victims of these campaigns are voters and political parties. Voters are targeted with misinformation intended to influence their voting decisions or discourage participation in the electoral process. Political parties face reputational damage as false information spreads about their candidates or platforms, potentially impacting their credibility with the public.

Emerging threats for 2024-2025 include the use of generative AI for disinformation. This technology allows adversaries to create highly convincing deep fakes and misinformation that are difficult to distinguish from genuine content [14], [28]. The communications security establishment (CSE) has assessed that foreign adversaries are likely to employ generative AI to target Canada's federal election in the coming years. Additionally, increased attribution difficulty poses a significant challenge; advanced obfuscation techniques make it harder to trace cyber threat activities back to specific state-sponsored actors. In 2022, 85% of reported cyber threat activities were unattributed [9].

*B. Attacks on election support (ES) infrastructure*

Attacks on election support (ES) infrastructure involve cyber threats targeting the systems and networks that facilitate the electoral process. These attacks can disrupt election operations, steal sensitive information, and spread disinformation, ultimately undermining public confidence in election results

*Adversary*: The primary adversaries include sophisticated malicious hackers and state-sponsored actors from foreign governments. These entities possess advanced skills and resources, allowing them to conduct operations with precision and persistence targeting specific infrastructure. State actors, particularly from nations like Russia and China, are known for leveraging cyber capabilities to influence political outcomes and erode trust in democratic institutions [9]. The operations often have long-term strategic objectives, destabilizing political environments and gathering intelligence.

*Capability*: Adversaries employ a suite of advanced tools and techniques for specific targets with precision. Malware is used strategically to infiltrate systems and exfiltrate data without detection. Phishing kits have evolved to include highly sophisticated social engineering tactics that exploit human vulnerabilities, targeting election officials and voters with tailored attacks. Additionally, attackers utilize network scanners to map out infrastructure vulnerabilities, identifying weak points for exploitation. Advanced password-cracking algorithms are used to bypass authentication mechanisms, while man-in-the-middle attacks intercept and manipulate encrypted communications, allowing adversaries to siphon or alter sensitive data. Distributed denial of service (DDoS) attacks has also become more precise, leveraging botnets to disrupt critical election-related services at critical moments.

*Infrastructure*: The infrastructure at risk extends beyond traditional voting systems to include interconnected digital platforms that support the electoral process. While Canada uses paper ballots for federal elections, electronic systems in certain regions are susceptible to cyber manipulation. Government databases containing voter registration and election official information are high-value targets due to the sensitive nature of the data they hold. Campaign websites, critical for political communication and voter engagement, face threats of defacement or service disruption, which can impede campaign operations.

*Victims*: These cyber-attacks impact both voters and the election commission. Voters risk exposure of personal data and potential manipulation of their electoral participation through misinformation campaigns. The election commission faces operational risks that could compromise its ability to ensure a fair and transparent electoral process.

Emerging threats include targeting cloud-based systems used by political parties and electoral systems, such as voter databases, campaign management platforms, and election infrastructure. As cloud services become integral to modern electoral systems, these attacks can involve data breaches, ransomware, and manipulation of sensitive information, potentially compromising the integrity of elections. There has been a notable 75% increase in cloud environment intrusions over the past year, reflecting adversaries' focus on exploiting vulnerabilities in cloud-based systems that support election infrastructure [29]. Additionally, malware-free attacks have surged from 40% to 75% between 2019 and 2023 [29]. These sophisticated attacks bypass traditional detection methods by using tools already present within target environments, making them particularly challenging to identify and mitigate.

## C. Cyber Espionage

Cyber espionage involves unauthorized access to information in cyberspace for political, or economic purposes. It is often conducted to gain a competitive advantage by infiltrating networks and extracting sensitive data.

*Adversary*: The primary adversaries involved in cyber espionage include foreign intelligence services and sophisticated cybercriminal groups. State-sponsored actors, particularly from nations with strategic interests in Canada, seek to gain a competitive advantage by infiltrating networks associated with Elections Canada, political party websites, and public servant communications [9]. These actors aim to gather intelligence that can influence political outcomes or be sold to interested parties. Sophisticated cybercriminal groups, often operating with the tacit approval or direct support of state entities, engage in similar activities for financial gain.

*Capability*: Advanced persistent threats (APTs) are a significant concern, characterized by their stealthy and continuous presence within target network [29]. These groups often utilize spear-phishing attacks, which are highly targeted and personalized email campaigns designed to gain unauthorized access to sensitive information. In addition to these tactics, APTs leverage zero-day exploits, allowing them to take advantage of previously unknown vulnerabilities in software systems before they can be patched. Once inside a network, they employ sophisticated data exfiltration techniques to covertly transfer sensitive information from compromised systems, ensuring that valuable intelligence is extracted without detection.

*Infrastructure*: The infrastructure at risk includes campaign networks, personal devices, cloud services, and email servers [30]. Campaign networks involve the IT systems utilized by political parties and candidates to store strategic information and communications, making them vital targets for attackers. Personal devices—such as smartphones, tablets, and laptops—belonging to key political figures are particularly vulnerable, serving as potential entry points for espionage. Additionally, cloud services used for storing campaign data and communications introduce further risks due to their accessibility from multiple locations and devices. Finally, email servers containing sensitive correspondence and strategic information are prime targets for data breaches.

*Victims*: The victims of cyber espionage efforts include political candidates, party officials, campaign staff, and volunteers. Political candidates are high-value targets due to their access to strategic campaign information and their influence over public opinion. Party officials also hold sensitive campaign strategies and donor information that adversaries aim to exploit. While campaign staff may often be less protected than senior officials, they still possess valuable information that can be leveraged by attackers. Additionally, volunteers can represent weak links in the security chain, as their varying levels of cybersecurity awareness and access to campaign operations make them vulnerable to exploitation.

Emerging threats for 2024-2025 include increasingly sophisticated APTs that employ zero-day exploits and long-term infiltration techniques. These advanced tactics have led to a 76% increase in the number of victims named on eCrime leak sites [29]. Additionally, the potential threat posed by quantum computing is a growing concern; as this technology advances, it could potentially break current encryption methods, compromising the confidentiality of sensitive campaign and party data.

## IV. Risk Assessment

In assessing the risk levels for the upcoming Canadian federal election, we consider three primary cyber threats: Misinformation, Disinformation, and Malinformation (MDM) campaigns; attacks on election support infrastructure; and cyber espionage targeting political parties and candidates. Each threat is evaluated based on its level of impact and probability, leading to an overall priority level.

### A. Misinformation, Disinformation, Malinformation (MDM)

Misinformation, disinformation, and malinformation (MDM) campaigns are assessed as the highest risk due to their significant impact and likelihood of occurrence. T hese campaigns profoundly affect reputation, public trust, political stability, and election outcomes. They receive a high impact score because the spread of disinformation can alter public discourse on a large scale, influencing voter behavior and fostering societal divisions [31]. This division can lead to the destabilization of political environments and significantly undermine public confidence in democratic institutions and processes [32], [33]. By spreading false information, MDM campaigns can manipulate voter perceptions and decisions, further complicating the electoral landscape.

The probability of these campaigns is also high due to the ease with which disinformation can be disseminated via digital platforms. The use of generative AI lowers the barrier to entry for disinformation campaigns, allowing malicious actors to produce high-quality false information at minimal cost. These tools enable the mass production of disinformation that can be easily spread through social media algorithms, targeting specific demographics with tailored content. Social media algorithms often amplify sensational content, increasing the reach of false narratives. These content can manipulate public opinion, demobilize voters, and increase societal divisions [32], [34]. An example of this can be seen in the 2021 Canadian federal election when disinformation campaigns targeted the conservative party of Canada (CPC) and MP Kenny Chiu [10]. Chiu had introduced a bill to create a foreign influence registry, a measure aimed at countering foreign interference. However, disinformation spread on Chinese-language social media, falsely claiming that the bill was anti-Chinese and that Chiu and the CPC were targeting the Chinese community. This campaign of disinformation may have impacted Chiu's re-election bid, and demonstrated how easily disinformation can be weaponized to influence electoral outcomes and sow divisions within communities.

### B. Attacks on election support (ES) infrastructure

Attacks on election support infrastructure can lead to severe service disruptions, data compromise, and diminished public confidence. The impact is significant because such disruptions can paralyze electoral operations and undermine public trust [35]. While Canada's use of paper ballots mitigates some risks associated with electronic voting systems, other elements such as voter registration databases and election management systems are still susceptible to cyber threats [36]. Disruptions in these areas can delay results and undermine public confidence in the electoral process. Additionally, the increase in cloud-based attacks poses an enhanced risk, as many political parties and electoral systems now rely on cloud storage for sensitive data, campaign web hosting, and other services. Compromises in these areas can have a substantial impact on the integrity and reliability of the electoral process.

Despite robust security measures, the probability of cyberattacks targeting critical infrastructure remains moderate to high, primarily due to the evolving sophistication of cybercriminals and state-sponsored actors. A notable example occurred in 2023 when Prime Minister Justin Trudeau's website was taken offline by a distributed denial-of-service (DDoS) attack. Although the attack did not directly affect election infrastructure, it illustrated how easily critical systems can be disrupted, potentially jeopardizing digital infrastructure and essential services.

### C. Cyber Espionage

Cyber espionage results in strategic data loss, reputational damage, and national security concerns. These activities tend to affect specific targets rather than the broader public, but it can still compromise sensitive campaign information, influence strategies, and expose vulnerabilities within political entities [9], [35]. Although it ranks third in terms of overall risk due to its targeted nature, the impact remains moderately high since stolen information can be strategically leveraged against political groups.

The probability of cyber espionage is also moderately high, as adversaries increasingly employ advanced techniques such as APTs and zero-day exploits. These tactics are typically utilized by well-resourced actors targeting key figures within political parties. A notable example occurred during the 2015 federal election when the Liberal Party of Canada was reportedly subjected to a cyber espionage operation. Foreign actors were suspected of attempting to steal sensitive campaign data, including internal communications and strategy documents, with the intent to disrupt or manipulate the electoral process. Although the exact details of the breach were not fully disclosed, this incident highlighted how cyber espionage can be leveraged to gain a strategic advantage, particularly by exploiting sensitive information from political parties during elections. Figure 2 illustrates the threat assessment of the three major threat.

### V. MITIGATION STRATEGIES AND RECOMMENDATIONS

#### A. Misinformation, Disinformation, Malinformation (MDM)

To effectively combat MDM in Canada, a multi-layered strategy is essential. One crucial component is enhancing digital literacy and public awareness. Nationwide digital literacy programs should be established to empower citizens to critically evaluate information sources [24]. These initiatives could include workshops, online courses, and public service announcements aimed at teaching individuals how to discern credible information from falsehoods [35]. It is important to ensure these programs are accessible to all, including marginalized communities, by leveraging community organizations such as public libraries and social service agencies [24], [37]. Additionally, public awareness campaigns can play a significant role in educating the public on recognizing misinformation and understanding the importance of credible sources, drawing on resources from organizations like MediaSmarts, which has been developing digital media literacy programs since 1996.

Another key element is the establishment of a dedicated task force to monitor and respond rapidly to disinformation campaigns on social media platforms [38]. This could include use advanced AI tools to detect patterns indicative of misinformation, enabling quick identification and response. Collaborating with social media companies is also vital to ensure the swift removal of harmful content and to enhance the platforms' own monitoring capabilities. By working closely with these companies, the task force can help create a more robust defense against the spread of false information. These partnerships would also facilitate the sharing of threat intelligence between platforms and government agencies, enhancing the overall response to MDM threats.

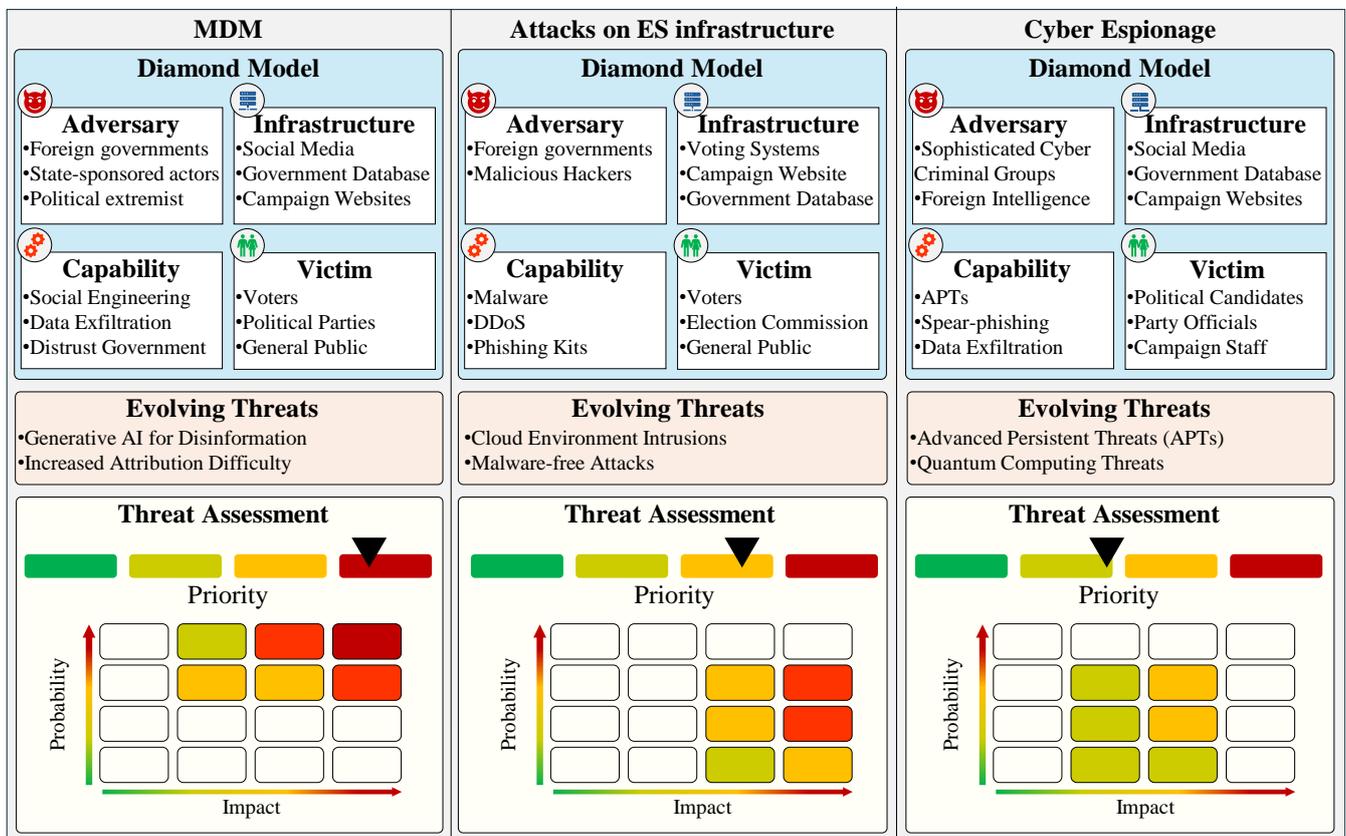

Fig.2. Diamond model analysis and threat assessment of the major cyber threats for the upcoming Canadian election.

*B. Attacks on election support (ES) infrastructure*

Securing election support infrastructure is crucial to maintaining the integrity and reliability of the electoral process. A comprehensive approach that prioritizes security can prevent disruptions and protect against cyber threats. This include a defense-in-depth (DiD) strategy which is essential for safeguarding election systems [39]. The strategy involves implementing a layered security model that includes firewalls, intrusion detection systems, and regular software updates. This ensures that if one layer is breached, additional defenses remain intact. Additionally, network segmentation should be utilized to isolate election systems from other government networks, minimizing the risk of lateral movement by attackers [39]. Furthermore, strict access controls must be enforced, employing multi-factor authentication (MFA) for all election-related systems to prevent unauthorized access [39].

Adopting a zero trust architecture further enhances security by continuously verifying access to systems, thereby reducing the risk of unauthorized entry. This approach requires strict identity verification processes and constant monitoring of user activities within election systems. Advanced intrusion detection systems (IDS) and intrusion prevention systems (IPS) should be deployed to monitor network traffic for signs of malicious activity [40]. Additionally, utilizing threat intelligence feeds can help keep security measures updated against emerging threats targeting election infrastructure.

Conducting Regular Security Assessments is vital for identifying vulnerabilities before they can be exploited by attackers. Frequent security audits and penetration testing should be carried out by independent cybersecurity experts to ensure objectivity and thoroughness. These assessments help uncover potential weaknesses in the system and provide insights into necessary improvements. Finally, incident response planning (IRP) is critical for effectively addressing cyber incidents [39]. Developing comprehensive incident response plans that outline roles, responsibilities, and communication protocols is necessary for coordinating efforts during an attack. Regular drills and simulations should be conducted to ensure readiness among election officials, enabling them to respond swiftly and effectively in the event of a cyber incident.

*C. Cyber Espionage*

One effective approach to protect against cyber espionage is to employ advanced threat detection technologies. Utilizing AI-driven analytics can help identify unusual patterns that may indicate espionage activities. These technologies are particularly useful in detecting APTs that might otherwise go unnoticed. Implementing endpoint detection and response (EDR) solutions can also detect APT activities early in their lifecycle, allowing for prompt intervention [20]. As election systems increasingly transition to cloud environments, robust cloud security measures, including identity threat detection and response (ITDR) systems, are essential to safeguard data integrity. Additionally, regularly updating software and applying security patches can mitigate zero-day vulnerabilities.

Encryption and data protection are crucial elements of a comprehensive cybersecurity strategy. It is essential to ensure that all sensitive data is encrypted both in transit and at rest, using strong cryptographic standards. To prepare for future threats posed by quantum computing, adopting post-quantum cryptographic techniques is advisable. In addition to these technical measures, targeted cybersecurity training should be provided for political candidates, party officials, and staff, with a focus on phishing awareness and secure communication practices. Furthermore, employing secure cloud storage solutions with robust access controls will help protect campaign data from unauthorized access [40].

Implementing comprehensive training programs is another key strategy in defending against cyber espionage. Regular cybersecurity training for political parties and candidates can help them recognize phishing attempts and secure their personal devices [20], [41]. Training should cover secure communication practices to protect sensitive information and include guidelines for securing personal devices used in campaign activities. Establishing a schedule for ongoing training updates ensures that staff and candidates remain informed about new threats as they emerge.

*D. General Recommendations*

To enhance cybersecurity measures against cyber threats, establishing robust information-sharing networks is vital. These networks foster collaboration among government agencies, law enforcement, cybersecurity experts, and international partners, such as the Five Eyes alliance, which includes the United States, United Kingdom, Canada, Australia, and New Zealand. By sharing intelligence and resources, these entities can significantly improve their threat detection and response capabilities, ensuring quicker and more effective action against potential cyber threats [35].

Additionally, engaging in public-private partnerships (PPPs) and strengthening regulatory frameworks are crucial strategies [42]. Collaborating with private sector experts allows governments to leverage cutting-edge cybersecurity technologies and innovative solutions. For instance, initiatives like the U.S. Cybersecurity and Infrastructure Security Agency's Joint Cyber Defense Collaborative demonstrate the effectiveness of such partnerships. Meanwhile, enhancing frameworks like the critical cyber systems protection act (CCSPA) ensures consistent cybersecurity standards across sectors, safeguarding critical infrastructure and election integrity through mandated cybersecurity plans and incident reporting requirements.

## VI. CONCLUSION

To safeguard the integrity of the upcoming Canadian federal election, a proactive approach is essential to address the major emerging cyber threats. This article identifies three primary threats: misinformation, disinformation, and malinformation (MDM) campaigns; attacks on election support infrastructure; and cyber espionage. It then conducts a thorough threat assessment of these risks, which pose challenges to election integrity, public trust, and national security. To mitigate these threats, the article recommends several strategies: enhancing digital literacy to empower citizens against misinformation, implementing robust cybersecurity measures to protect election infrastructure, and fostering collaboration between government and private sectors to strengthen defenses. By adopting these strategies, Canada can safeguard its democratic processes against increasingly sophisticated cyber adversaries, ensuring a secure and transparent electoral process.

ACKNOWLEDGEMENT